\documentclass[allclo]{FBSart}
\usepackage{amsfonts}
\usepackage{amssymb}
\newif\ifpdf
        \ifx\pdfoutput\undefined
        \pdffalse 
        \else
        \pdfoutput=1 
        \pdftrue
        \fi
\ifpdf
        \usepackage[pdftex]{graphicx}
        \DeclareGraphicsExtensions{.pdf, .jpg}
\else
        \usepackage[dvips]{graphicx}
                \DeclareGraphicsExtensions{.eps, .jpg}
\fi

\title{Structure of exotic three-body systems}
\author{T.~Frederico$^1$\thanks{\textit{E-mail address:}
 tobias@ita.br}, M.T.~Yamashita$^2$, A.~Delfino$^3$, Lauro~Tomio$^4$}
\institute{$^1$Departamento de F\'\i sica, Instituto Tecnol\'ogico
de Aeron\'autica, Centro T\'ecnico Aeroespacial, 12.228-900 S\~ao
Jos\'e dos Campos, Brasil. \\ $^2$Unidade Diferenciada de Itapeva,
Universidade Estadual Paulista, 18.409-010 Itapeva,
 Brasil. \\
$^3$ Instituto de F\'\i sica, Universidade Federal Fluminense,
24.210-900 Niter\'oi, Brasil. \\
 $^4$Instituto de F\'\i sica Te\'orica, Universidade
Estadual Paulista,  01405-900 S\~{a}o Paulo,  Brasil }

\runningauthor{ T.~Frederico, M.T.~Yamashita, A.~Delfino,
Lauro~Tomio} \runningtitle{ Structure of exotic three-body systems
}  
\begin{document}

\maketitle
\begin{abstract}
The classification of large halos formed by two identical
particles and a core is systematically addressed according to
interparticle distances. The root-mean-square distances between
the constituents are described by universal scaling functions
obtained from a renormalized zero-range model. Applications for
halo nuclei, $^{11}$Li and $^{14}$Be, and for atomic $^4$He$_3$
are briefly discussed. The generalization to four-body systems is
proposed.

\end{abstract}

\section{Introduction}
The properties of large three-body systems present universal
behavior as the Thomas-Efimov limit is
approached~\cite{jensenrmp}. In this situation the sizes of the
two and three body systems are much larger than the interaction
range and the wave function almost everywhere is an eigenstate of
the free Hamiltonian. Much of the physics of these systems can be
studied using a zero-range interaction effective in s-wave. No
scale is involved in this drastic situation and the momentum space
Faddeev equations are scale invariant for vanishing two and
three-body energies. The scale invariance is broken by finite
values of two-body scattering lengths and three-body binding
energy, which are the scales of the corresponding three-body
systems~\cite{AdPRL95}. Dimensionless ratios of three-body
observables are then expressed by scaling functions which depends
only on ratios of two (bound or virtual) and three-body
energies~\cite{amorim1999}. The scaling functions can be obtained
from the solution of regularized Faddeev equations.

The sizes of halo three-body systems (where two particles are
identical) are functions of few physical scales. A classification
scheme for these systems~\cite{robi99,jensen03,raiosnpa} ordered
by their sizes is reviewed. For a given three-body binding energy
the most compact system is the Borromean (single pairs are
unbound) while the all-bound (all single pairs form bound states)
is the largest one. The neutron-neutron mean square radius in
examples of Borromean halo nuclei which are known
experimentally~\cite{marques} are briefly discussed. The extension
of these ideas to four-body systems is proposed.

\section{Scale invariance in three-boson systems}

The collapse of a three-boson system when the two-body interaction
range goes to zero (Thomas effect, see ~\cite{jensenrmp}) demands
one new physical scale to stabilizes the three-boson binding
energy ($E_3$).  The regulated trimer bound state integral
equation in units of $\hbar=m=1$ ($m$ is the boson mass) is
written as
\begin{eqnarray}
f(q)=2\pi\tau(E_3-\frac34q^2)\int_0^\infty k^2dk~f(k)
\int^1_{-1}dz\left[G_0(E_3)-G_0(-\mu_{(3)}^2) \right] ,
\label{trimer}
\end{eqnarray}
where $G^{-1}_0(E)=E-q^2-k^2-qkz$, and the renormalized two-boson
scattering amplitude is
$\tau^{-1}(x)=2\pi^{2}\left[a^{-1}-\sqrt{-x}\right]. $ Without
regularization, Eq.~(\ref{trimer}) is the Skorniakov and
Ter-Martirosian (SKTM) equation derived long ago~\cite{skt} which
is scale invariant when the scattering length $|a|$ tends to
infinity and $E_3=0$. The second term in Eq.~(\ref{trimer}) brings
the physical scale, $\mu_{(3)}$, to the three-boson
system~\cite{AdPRL95} avoiding the Thomas collapse.  The
equivalence of Thomas and Efimov effects is seen in units of
$\mu_{(3)}=1$, which means $|a|\mu_{(3)}\rightarrow\infty$ either
for $|a|\rightarrow\infty$ (Efimov limit) or for
 $\mu_{(3)}\rightarrow\infty$ (Thomas limit).

The sensitivity of  three-boson S-wave observables to the
short-range part of the interaction in weakly bound systems is
parameterized through the value of the trimer binding energy which
corresponds to the scale $\mu_{(3)}$.  Three-boson S-wave
observables are strongly correlated to the trimer energy  in a
general universal form~\cite{amorim1999,virtual}:
\begin{eqnarray}
{\cal O}_3(E, E_3, a)=|E_3|^\eta {\cal
F}_3\left(E/E_3,a\sqrt{|E_3|}\right) \ , \label{obs3}
\end{eqnarray}
where ${\cal O}_3$ can represent a scattering amplitude at an energy
$E$ or an excited trimer energy (the dependence on $E$ does not
appear in this case). The exponent $\eta$ gives the correct
dimension to ${\cal O}_3$. Eq.~(\ref{trimer}) is renormalization
group (RG) invariant with  its kernel being a solution of a
Callan-Symanzik differential equation as function of a sliding
$\mu_{(3)}$. In that way $E_3$ and three-body observables are
independent of the subtraction point (see ref.~\cite{virtual} for
a discussion of the RG invariance in three-body systems).

\section{Classification of three-body halos and universal scalings}

The sizes of halo three-body systems (with two identical particles
($\alpha$) and a distinct one ($\beta$)) are
functions of few physical scales. A classification scheme for
these systems as Borromean (single pairs are unbound), Tango (only
the  $\alpha\alpha$ pair form a bound state), Samba (only the
$\alpha\beta$ pair is bound) and all-bound (all single pairs form
bound states), ordered by their sizes is discussed below.

The typical lengths of the three-body halo systems are given by
scaling functions for the mean-square separation distances written
according to Eq.(\ref{obs3}). The scaling functions for these
radii are:
\begin{eqnarray}
&&\sqrt{\langle r^2_{\alpha\gamma}\rangle |E_{3}|} ={\cal
R}_{n\gamma}\left(\pm \sqrt{E_{\alpha\alpha}/E_{3}},
\pm\sqrt{E_{\alpha\beta}/E_{3}}, A\right) , \label{rag}
\\
&&\sqrt{\langle r^2_{\gamma}\rangle |E_{3}|}= {\cal
R}_{\gamma}^{cm}\left(\pm \sqrt{E_{\alpha\alpha}/E_{3}},
\pm\sqrt{E_{\alpha\beta}/E_{3}}, A\right) , \label{ragcm}
\end{eqnarray}
where $\langle r^2_{\alpha\gamma}\rangle$ and $\langle
r^2_{\gamma}\rangle$ are, respectively, the mean square relative
and center of mass distances. The mass ratio is
$A=m_\beta/m_\alpha$ and $\gamma$ is $\alpha$ or $\beta$. The + or
- signs represent bound or virtual two-body subsystems,
respectively. $E_{\alpha\alpha}$ and $E_{\alpha\beta}$ are the
$\alpha\alpha$ and $\alpha\beta$ two-body energies.

For a given energy $E_3$ and identical particles, the effective
interaction in Eq.~(\ref{trimer}) has a weaker strength for $a<0$
(virtual two-body system) than for $a>0$ (bound two-body system).
Therefore, for a Borromean trimer Eq.~(\ref{trimer}) should have a
larger value of $\mu_3$ than the corresponding one for an
all-bound system, in order to keep the binding fixed with a weaker
interaction. The spectator function, $f(q)$, extends to large
momentum for a Borromean system. The trimer wave function for zero
total angular momentum is
\begin{eqnarray}
\Psi(q,p)=\frac{f(|\vec q|)+f(|\vec p+\frac12 \vec q|)+f(|\vec
p-\frac12 \vec q|)}{E_3-\frac34q^2-p^2}~, \label{wf}
\end{eqnarray}
which implies in a more compact spatial configuration for a
Borromean trimer in comparison to the all-bound one. The Jacobi
relative momenta are $\vec p$ for the pair and $\vec q$ for the
spectator particle. In terms of the scaling functions the radii
come as,
\begin{eqnarray}
{\cal R}_{\alpha\alpha}\left(
-\sqrt{{E_{\alpha\alpha}}/{E_{3}}}\right) < {\cal
R}_{\alpha\alpha}\left(\sqrt{{E_{\alpha\alpha}}/{E_{3}}}\right) ,
\label{ragt}
\end{eqnarray}
and the separation distances obeys $\sqrt{\langle
r^2_{\alpha\alpha}\rangle |E_{3}|}_B<\sqrt{\langle
r^2_{\alpha\alpha}\rangle |E_{3}|}_A $, where the labels $B$ and
$A$ correspond to Borromean and All-bound systems, respectively.
For the same reasons that led to Eq.~(\ref{ragt}), it is also
valid that $\sqrt{\langle r^2_{\alpha}\rangle
|E_{3}|}_B<\sqrt{\langle r^2_{\alpha}\rangle |E_{3}|}_A $. The
zero-range model applied to atomic $^4$He$_3$ provides a
qualitative understanding of the radii results of realistic
calculations for the ground and excited states~\cite{barletta}
with an estimation of $\sqrt{\langle
r^2_{\alpha\alpha}\rangle}={\cal
C}\sqrt{\hbar^2/[m_\alpha(|E_3-E_{\alpha\alpha}|)]}$ with
$0.6<{\cal C}<1$~\cite{raiosnpa}. Also the excited trimer state
energy of $^4$He$_3$, $E^*_3$, has a scaling behavior written as
\begin{eqnarray}
\sqrt{|E^*_3-E_{\alpha\alpha}|}=\sqrt{|E_3|}~{\cal
E}\left(\pm\sqrt{E_{\alpha\alpha}/E_3}\right) \ , \label{e3*}
\end{eqnarray}
which is consistent with results from realistic calculations (see
~\cite{amorim1999,virtual} and references therein). The threshold
for the appearance of an excited Efimov trimer state from the
second energy sheet is $|E_3|=6.9/a^2$~\cite{virtual,braaten} in
units of ($\hbar=m_\alpha=1$). The scaling behavior  was extended
to the complex energy plane and for three-boson Borromean systems,
the excited Efimov state turns into a resonance when the virtual
two-boson virtual state energy is decreased~\cite{resonance}.

The generalization of the reasonings leading to Eq.~(\ref{ragt})
to the $\alpha\alpha\beta$ system gives the qualitative
classification of the different three-body systems in respect to
sizes. The effective interaction is weaker when a pair has a
virtual state than when the pair is bound, and the three-body
system has to shrink to keep the binding energy unchanged if a
pair which is bound turns to be virtual. Therefore it is
reasonable to expect that
\begin{eqnarray}
{\cal R}_{\alpha\gamma}\left(-\sqrt{{E_{\alpha\alpha}}/{E_{3}}},
-\sqrt{{E_{\alpha\beta}}/{E_{3}}}, A\right)<{\cal
R}_{\alpha\gamma}\left(\sqrt{{E_{\alpha\alpha}}/{E_{3}}},
-\sqrt{{E_{\alpha\beta}}/{E_{3}}}, A\right)\nonumber \\ <{\cal
R}_{\alpha\gamma}\left(-\sqrt{{E_{\alpha\alpha}}/{E_{3}}},
\sqrt{{E_{\alpha\beta}}/{E_{3}}}, A\right)<{\cal
R}_{\alpha\gamma}\left(\sqrt{{E_{\alpha\alpha}}/{E_{3}}},
\sqrt{{E_{\alpha\beta}}/{E_{3}}}, A\right)~,
\end{eqnarray}
which was checked numerically~\cite{raiosnpa}. An analogous
relation is valid for ${\cal R}_{\gamma}^{cm}$.  The dimensionless
products (we are using units of $\hbar=m_\alpha=1$) $\sqrt{\langle
r^2_{\alpha\beta} \rangle |E_3|}$ and $\sqrt{\langle
r^2_{\alpha\alpha} \rangle |E_3|}$ increase from {\it Borromean},
{\it Tango}, {\it Samba} and to {\it All-Bound} configurations,
systematizing  the classification scheme proposed in
Ref.~\cite{jensen03} for weakly bound three-body systems.

A three-body model applied to light exotic nuclei~\cite{raiosnpa}
compares qualitatively well with the existent experimental data
for the neutron-neutron separation distance in the neutron-halo of
$^{11}$Li and $^{14}$Be~\cite{marques}.  Therefore, the neutrons
of the halo have a large probability to be found outside the
interaction range and the low-energy properties of the halo are,
to a large extend, universal as long as few physical input scales
are fixed in  the model. An insight into the structure of halo
nuclei can be found even considering the limitations of the model.
The finite size of the core and consequently the
antisymmetrization of the total nuclear wave function, are both
missing in this simplified description. However, in examples where
the neutrons in the halo tend to be more and more weakly bound
with virtual or bound subsystems near the scattering threshold,
the scaling relations apply for the halo properties, as the above
limitations are less important.

\section{Classification of four-body halos and universal scalings}

The four-boson system has two Faddeev-Yakubovsky (FY) independent
amplitudes and within a zero-range model, they are reduced to
spectator amplitudes depending on two Jacobi momenta. The two
spectator FY reduced amplitudes satisfy a coupled set of integral
equations generalizing the SKTM equation for three bosons. The set
of coupled integral equations needs regularization, and one
recognizes that the resolvent of the immersed three-boson subsystem
carries the scale $\mu_{(3)}$. Other terms are present and
require regularization. We introduce a scale $\mu_{(4)}$ such
that the four-body free Green's function $G_0(E_4)$ are
substituted by $G_0(E_4)-G_0(-\mu_{(4)}^2)$~\cite{4body} in a
direct generalization of Eq.~(\ref{trimer}) as suggested by
~\cite{AdPRL95}.

The momentum scales in the FY equations for the reduced amplitudes
are only $a^{-1}$, $\mu_{(3)}$ and $\mu_{(4)}$. In this case, the
tetramer binding energy depends on the momentum scales as
\begin{eqnarray}
E_4=\mu^2_{(3)}~{\cal \overline{E}
}_4\left(\mu_{(4)}/\mu_{(3)},a\mu_{(3)}\right). \label{sca4b}
\end{eqnarray}
For $a=\infty$ the trimer binding energy from the solution of
Eq.~(\ref{trimer}) is $E_3 =-0.0093\mu^2_{(3)}$~\cite{virtual},
which simplifies Eq.~(\ref{sca4b}) remaining only the dependence
on the ratio $\mu_{(4)}/\mu_{(3)}$ in $E_4$:
\begin{eqnarray}
E_4=E_3~{\cal E }_4\left(\mu_{(4)}/\mu_{(3)}\right).
\label{sca4b1}
\end{eqnarray}
If ${\cal {E} }_4$ is independent on the regularization scale
$\mu_{(4)}$ for $\mu_{(4)}/\mu_{(3)}>>1$ the four-body scale is
not important. We solved numerically the FY equations up to large
values of $\mu_{(4)}/\mu_{(3)}\sim 20$~\cite{4body}.

The tetramer ground state binding energy was calculated for
different values of the ratio $\mu_{(4)}/\mu_{(3)}$ with
$a=\infty$. The ratio $E_4/E_3$ depends strongly
$\mu_{(4)}/\mu_{(3)}$. For equal three and four-body scales, i.e.,
$\mu_{(4)}/\mu_{(3)}=1$, the tetramer binding energy is
$E_4=5~E_3$, agreeing with the angular coefficient of the Tjon
line~\cite{tjon}, $E_4=4.72(E_3+2.48)$~MeV ($E_4$ is the $^4$He
energy and $E_3$ the triton one). Also, a recent
calculation~\cite{platter} of the four-boson system with a
two-body zero-range force and a repulsive three-body potential to
stabilize the trimer against collapse, $E_4$ scales as $\sim
5~E_3$. Our result for $\mu_{(4)}=\mu_{(3)}$ agrees with both.
However, for $\mu_{(4)}/\mu_{(3)}=20$ we found that the ratio
between the tetramer and trimer energies is about 78,  indicating
the independent effect of the four-body scale.

The present results suggest that the general scaling of  S-wave
three-boson observables with the physical scales,
Eq.~(\ref{obs3}), may be generalized to four-boson S-wave
observables. The effect of the short-range dynamics in an
observable comes through the values of the scattering length,
trimer and tetramer binding energies, associated with $\mu_{(3)}$
and $\mu_{(4)}$, respectively. In this case, a S-wave four-boson
observable will be strongly correlated to $a$, $E_3$ and $E_4$:
\begin{eqnarray}
{\cal O}_4(E, E_4,E_3, a)=|E_4|^\eta {\cal
F}_4\left({E}/{E_4},{E_3}/{E_4},a\sqrt{|E_4|}\right) \ ,
\label{obs4}
\end{eqnarray}
where ${\cal O}_4 $ represents either a scattering amplitude at
energy $E$, or an excited tetramer energy or some observable
related to the tetramer (the dependence on $E$ does not appear in
these cases). The exponent $\eta$ gives the correct dimension to
${\cal O}_4 $. For sizes one could think that the relation
\begin{eqnarray}
{\cal R}_{\alpha\alpha}\left(\sqrt{E_3/E_4},
-\sqrt{E_{\alpha\alpha}/E_4}\right) < {\cal
R}_{\alpha\alpha}\left(\sqrt{E_3/E_4},\sqrt{E_{\alpha\alpha}/E_4}\right)
, \label{ragt4}
\end{eqnarray}
and the analogous for the distances of the particles to the center
of mass would be valid. This indicates that is possible to
envisage a generalized  classification scheme based on sizes
including weakly-bound four-body systems.

\section{Outlook and Conclusions}
The classification scheme of large halos formed by two identical
particles and a core is reviewed and addressed systematically
according to their sizes. The root-mean-square distances between
the constituents are described by universal scaling functions. For
a given three-body system and total energy, the Borromean
configuration is the most compact. Applications to halo nuclei,
$^{11}$Li and $^{14}$Be, and for atomic $^4$He$_3$ were briefly
discussed.

The generalization of these concepts to four-body systems is
proposed. We have shown that for a zero-range two-body interaction
with an infinite scattering length and a fixed trimer ground state
binding energy, a four boson momentum scale is evidenced in the
calculation of tetramer binding energies in three-dimensions. The
intensity of the effective interaction that composes the kernel of
the reduced FY spectator equations depends on the dimer energy and
trimer, therefore it is reasonable that a tetramer becomes more
compact for a given four-body energy if the two and three-body
binding are decreased. This effect indicates that it may be
possible to generalize the  classification scheme of weakly-bound
three-body systems to four-body systems.

\begin{acknowledge}
We thank the Brazilian funding agencies FAPESP and CNPq.
\end{acknowledge}

\end{document}